\documentstyle[12pt]{article}

\def\be{\begin{equation}}
\def\ee{\end{equation}}
\def\ba{\begin{array}{c}}
\def\ea{\end{array}}
\def\p{\partial}
\def\ben{$$}
\def\een{$$}
\begin{document}

\titlepage
\vspace*{4cm}

 \begin{center}
{\Large \bf Solvable analogue of $V(x)=i\,x^3$
 }

\end{center}

\vspace{5mm}

 \begin{center}

Miloslav Znojil
\vspace{3mm}

\'{U}stav jadern\'e fyziky AV \v{C}R, 250 68 \v{R}e\v{z},
Czech Republic\\

e-mail: znojil@ujf.cas.cz

\end{center}

\vspace{5mm}

\section*{Abstract}
We prove that the purely imaginary square well generates an
infinite number of bound states with real energies. In the
strong-coupling limit, our exact ${\cal PT}$ symmetric solutions
coincide, utterly unexpectedly, with their textbook, well known
Hermitian predecessors.

 \vspace{9mm}

\noindent PACS 03.65.Bz, 03.65.Ge

 \vspace{9mm}

 \begin{center}
 {\small \today, sho.tex file}
 \end{center}

\newpage

\section{Introduction}

An interest in the imaginary cubic anharmonic oscillators dates
back to their perturbation analysis by Caliceti et al
\cite{Caliceti}. The simplified homework example with the mere
two-term non-Hermitian Hamiltonian
 \ben
 H_{BZ}=p^2+i\,x^3
  \een
has been proposed by D. Bessis and J. Zinn-Justin who had in mind
its possible applicability in the context of statistical physics
\cite{DB}. The example has been re-vitalized by C. Bender et al
due to its further possible methodical relevance in the
relativistic quantum field theory \cite{BM}.

The apparent reality of the spectrum of energies $E_{BZ}$ proved
quite puzzling and inspired a conjecture of the existence of the
whole ``modified quantum mechanics" paying attention to similar
Hamiltonians \cite{BBjmp}. This conjecture opened many new and
interesting questions when it replaced the current Hermiticity of
Hamiltonians by a weaker condition of their commutativity with a
product ${\cal PT}$ of the spatial parity ${\cal P}$ and the
complex conjugation ${\cal T}$ where the latter factor is to be
understood as a one-dimensional version of the operator of time
reversal.

The recent discussion between C. Bender and A. Mezincescu
\cite{Mezincescu} pointed out that one of the key problems of the
new studies lies in the ambiguity of the spectrum which depends
quite crucially on our choice of the boundary conditions which can
be, in general, complexified \cite{BT}. The fragile character of
the reality of energies has been confirmed by the WKB and
perturbative studies \cite{Trinh} and by the quasi-exact and exact
models \cite{QES} where the admissible unavoided level crossings
\cite{ptho} prove sometimes followed by the spontaneous breakdown
of the ${\cal PT}$ symmetry \cite{quartic}.

In such a context we intend to propose an extremely elementary
${\cal PT}$ symmetric model which would replace the BZ interaction
$ix^3$ (which admits just a numerical treatment) by its exactly
solvable square-well analogue.

\section{Model}

In a search for analogies between the solvable and unsolvable
models in one dimension, all the possible forms of a confining
well are often being approximated by the ordinary real and
symmetric square well
  \be
 V^{(SQW)}(x) =
 \left \{
 \begin{array}{ll}
  S^2,& x \in (-\infty,-\pi)\bigcup (\pi,\infty),\\
 0,
 &
   x \in (-\pi,\pi).
   \ea
   \right .
 \label{SQW}
  \ee
In this spirit we intend to replace here the above-mentioned
antisymmetric and imaginary homework potential $V_{BZ}(x)=i\,x^3 $
by its elementary square-well analogue
  \ben
 V^{(ISQW)}(x) =
 \left \{
 \begin{array}{ll}
  -i\,T^2,& x \in  (-\infty,-\pi),\\
 0,  &    x \in (-\pi,\pi),\\
  +i\,T^2,& x \in  (\pi,\infty).
   \ea
   \right .
  \een
Schr\"{o}dinger equation which appears in such a setting,
 \be
\left [ -\frac{\hbar}{2m}\,\frac{d}{dx^2}+V^{(ISQW)}(x) \right ]
\psi(x)=E\psi(x)
 \label{ISQW}
 \ee
will be complemented  here by the standard ${\rm L}^2(l\!\!R)$
boundary conditions
 \be
\psi(\pm \infty) = 0.
 \label{bc}
 \ee
The well known ${\cal PT}$ symmetric normalization convention will
be employed, with a free real parameter $G$ in the unbroken ${\cal
PT}$-symmetry requirements \cite{norma}
 \be
  \psi(0)=1,
 \ \ \ \ \ \ \ \p_x \psi(0)=i\,G.
  \label{nor}
 \ee
Putting $\hbar=2m=1$ and using the ansatz
  \be
 \psi(x) =
 \left \{
 \begin{array}{llc}
   \cos k\,x + B\,\sin k\, x,
    & x \in (0,\pi),& k^2 = E,\\
 (L + i\,N)\,\exp (-\sigma\,x)
  ,& x \in (\pi,\infty),&
\sigma^2=i\,T^2-k^2,
   \ea
   \right .
 \label{ansatz}
  \ee
we guarantee its full compatibility with the symmetry requirements
(\ref{nor}) by the choice of the purely imaginary constant $B
=i\,G/k$ in the wave functions (\ref{ansatz}).

\section{Matching conditions at $x=\pi$}

We may split $\sigma = p + i\,q$ in its real and imaginary part
with $p,q \geq 0$. This gives the rules $p^2+k^2=q^2$ and
$2pq=T^2$ as a consequence. They are easily re-parameterized in
terms of a single variable $ \alpha$,
 \be
p=q\,\cos \alpha, \ \ \ \ \ k =q\,\sin \alpha, \ \ \ \ \ q
=\frac{T}{\sqrt{2\cos \alpha}}, \ \ \ \ \ \  \alpha \in (0,
\pi/2).
 \label{param}
 \ee
In this language the standard matching at the point of
discontinuity is immediate,
 \ben
 \cos k\pi + B\, \sin k \pi =
 (L + i\,N)\,\exp (-\sigma\,\pi),
 \een
 \ben
 -\sin k\pi + B\, \cos k \pi =
 -\frac{\sigma}{k}(L + i\,N)\,\exp (-\sigma\,\pi).
 \een
After we abbreviate $\sigma/k=-\tan \Omega\pi $, we get an
elementary complex condition of the matching of logarithmic
derivatives at $x=\pi$,
 \be
 G = -i\,k\,\tan(k+\Omega)\pi.
 \label{secular}
 \ee
Its real part defines our first unknown parameter, $G =
G(\alpha)$. Due to our normalization conventions, the imaginary
part of the right-hand-side expression must vanish, ${\rm
Re}[\tan(k+\Omega)\pi]=0$. An elementary re-arrangement of such an
equation acquires the form of an elementary quadratic algebraic
equation for $X=\tan k\pi$. Its two explicit solutions read
 \be
X_1 = \frac{p+q}{k}, \ \ \ \ \ \ \ \ \ \ X_2 = \frac{p-q}{k}
 \label{sice}
 \ee
or, after all the insertions,
 \be
 \tan \left [
{\frac{ \pi T \sin \alpha^{(+)}}{\sqrt{2 \cos \alpha^{(+)}}
 }} \right ]=
 {\rm tan} \left [ \frac{\pi-\alpha^{(+)}}{2} \right ],
 \label{sicep}
 \ee
 \be
 \tan \left [
{\frac{ \pi T \sin \alpha^{(-)}}{\sqrt{2 \cos \alpha^{(-)}}
 }} \right ]=
 \tan \left [- \frac{\alpha^{(-)}}{2} \right ].
 \label{sicem}
 \ee
These equations specify, in implicit manner, the two respective
infinite series of the appropriately bounded real roots
$\alpha=\alpha^{(\pm)}_n \in (0, \pi/2)$.

\section{Energies}

Even before any numerical considerations we immediately see that
for $ \alpha \in (0, \pi/2)$ the left-hand-side arguments $[
\ldots ]$ in eqs. (\ref{sicep}) and (\ref{sicem}) run from zero to
infinity. Their tangens functions oscillate infinitely many times
from minus infinity to plus infinity. Within the same interval,
the limited variation of the argument $\alpha$ makes both the
eligible right-hand side functions monotonic, very smooth and
bounded, $ {\rm tan} [{(\pi-\alpha^{(+)})}/{2}] \in (1,\infty)$
and $ {\rm tan}[ {\alpha^{(-)}}/{2}] \in (0,1)$. {\it A priori}
this indicates that our roots $k=k(\alpha_n^{(\pm)})$ will all lie
within the fairly well determined intervals,
 \ben
 k_n^{(+)} \in
 \left ( n+\frac{1}{4},n+\frac{1}{2} \right ),
 \ \ \ \ \ \ \ \ n = 0, 1, \ldots,
 \een
 \ben
 k_m^{(-)} \in
 \left ( m+\frac{3}{4},m+{1} \right )
 \ \ \ \ \ \ \ \ m = 0, 1, \ldots.
 \een
After such an approximate localization of the roots, an unexpected
additional merit of our parametrization (\ref{param}) manifests
itself in an unambiguous removal of the tangens operators from
both eqs. (\ref{sicep}) and (\ref{sicem}). This gives the
following two relations,
 \ben
 k_n^{(+)}= n+\frac{1}{2}-\frac{\omega_{n}^{(+)}}{4}
 , \ \ \ \ \ \ \ \ \
 k_m^{(-)}= m+{1}-\frac{\omega_m^{(-)}}{4},
 \ \ \ \ \ \ \ \ \ \ \omega_n^{(\pm)}=
 \frac{2\alpha_n^{(\pm)}}{\pi}\  \in \ (0, 1).
 \een
After an elementary change of the notation with $\omega_n^{(+)}
=\omega_{2n}$ and $\omega_n^{(-)}=\omega_{2n+1}$, we may finally
combine the latter two rules in the single secular equation
 \be
\sin \left (
 \frac{\pi}{2}\omega_N \right )=
 \frac{2N+2-\omega_N}{4T}\cdot \sqrt{2 \cos \left (
 \frac{\pi}{2}\omega_N \right )}
 \ \ \ \ \ \ \ \ N = 0, 1, \ldots,
 \ \ \ \ \ \ \ \ \
 \label{enep}
 \ee
In a graphical interpretation this equation represents again an
intersection of a tangens-like curve with the infinite family of
parallel lines. This is illustrated in Figure~1. The equation
generates, therefore, an infinite number of the real roots
$\omega_N \in (0,1)$ at all the non-negative integers $N = 0, 1,
\ldots$.

\section{Wave functions in the weak coupling regime}

Equation (\ref{secular}) in combination with eqs. (\ref{sicep})
and (\ref{sicem}) determines the real parameter
 \be
G= G^{(\pm)}=-\frac{k^2}{q\pm p}
 \label{ecu}
 \ee
responsible for the behaviour of the wave functions near the
origin [remember that $B =iG/k$ in eq. (\ref{ansatz})]. For its
deeper analysis let us first introduce an auxiliary linear
function of $\omega$ and $N$,\
 \ben
  \sqrt{R(\omega_N,N)}=
 \frac{2N+2-\omega_N}{4T}\  \in \ \left (
 \frac{N+1/2}{2T}\ , \frac{N+1}{2T} \right )
 \een
and re-interpret our secular eq. (\ref{enep}) as an algebraic
quadratic equation with the unique positive solution,
 \be
 \cos \left (
 \frac{\pi}{2}\omega_N \right ) = \frac{1}{R(\omega_N,N)
 +\sqrt{R^2(\omega_N,N)+1}} \ .
 \label{secu}
 \ee
This is an amended implicit definition of the sequence $\omega_N$.
As long as the right hand side expression is very smooth and never
exceeds one, the latter formula re-verifies that the root
$\omega_N$ is always real and bounded as required.

In the domain of the large and almost constant $R \gg 1$ (i.e.,
for the small square-well height $T$ or at the higher
excitations), our new secular equation (\ref{secu}) gives a better
picture of our bound-state parameters $\omega_N= 1-\eta_N$ which
all lie very close to one. The estimate
 \ben
 \frac{\pi}{2}\,\eta_N = \arcsin \frac{1}{R+\sqrt{R^2+1}} \approx
 \frac{1}{2R} - \frac{5}{48\,R^3} + \ldots\
 \een
represents also a quickly convergent iterative algorithm for the
efficient numerical evaluation of the roots $\omega_N$. One can
conclude that in a way compatible with our {\it a priori}
expectations, the value of $p=p_N={\rm Re} \sigma \approx q/2R$ is
very close to zero and, as a consequence, the asymptotic decrease
of our wave functions remains slow. We have $q=q_N={\rm Im} \sigma
\approx k$ so that, asymptotically, our wave functions very much
resemble free waves $\exp( -i k x)$. In the light of eq.
(\ref{ecu}) we have also $\psi(x) \approx \exp( -i k x)$ near the
origin.

\section{Wave functions in the strong coupling regime}

For the models with a very small $R$ (i.e., for the low-lying
excitations in a deep well with $T \gg 1$) we get an alternative
estimate
 \ben
 \frac{\pi}{4}\,\omega_N = \arcsin
 \sqrt{
 \frac{1}{2}
 \left [ R- \left (
 \sqrt{1+R^2}-1
 \right )
 \right ]
 }
  \approx
 \frac{1}{2}\,R-\frac{1}{4}R^2 + \ldots\  \ll  \frac{\pi}{4}.
 \een
In the limit $R \to 0$ the present spectrum of energies moves
towards (and precisely coincides with) the well known levels of
the infinitely deep Hermitian square well of the same width $I =
(-\pi,\pi)$ (cf. eq. (\ref{SQW}) with $S \to \infty$). In this
sense, the ``complex-rotation" transition from the Hermitian well
$V^{(SQW)}(x)$ of eq. (\ref{SQW}) (with $S \gg 1$) to its present
non-Hermitian ${\cal PT}$ symmetric alternative $V^{(ISQW)}(x)$ of
eq. (\ref{ISQW}) (with $T \gg 1$) proves amazingly smooth.

The wave functions exhibit the similar tendency. In the outer
region, they are proportional to $ \exp ( -px)$ and decay very
quickly since $p = {\cal O}(R^{-1/2})$. The parameter $G^{(\pm)}$
becomes strongly superscript-dependent,
 \ben
 G^{(+)} = -\frac{k^2}{q+p} ={\cal O}(R^{3/2}),
 \ \ \ \ \ \
 G^{(-)} = -(q+p)={\cal O}(R^{-1/2}).
 \een
This means that in the interior domain of $x \in (-\pi,\pi)$, the
wave functions with the superscript $^{(+)}$ and $^{(-)}$ become
dominated by their spatially even and odd components $\cos kx$ and
$\sin kx$, respectively. In this sense, the superscript mimics (or
at least keeps the trace of) the quantum number of the slightly
broken spatial parity ${\cal P}$.

We can summarize that our present ${\cal PT}$ symmetric model is,
unexpectedly, quite robust. Almost irrespectively of the coupling
$T$, the spectrum is unbounded from above and remains constrained
by the inequalities
 \be
  \frac{(N+1/2)^2}{4}\ \leq \
E_N \  \leq \ \frac{ (N+1)^2}{4}\
  .
 \label{ene}
 \ee
The analogy between our exactly solvable square-well model and the
standard or ``paradigmatic" ${\cal PT}$ symmetric Hamiltonian
$H_{BZ}$ appears to be closer than expected.

\section{Outlook}

The exact solvability of our present purely imaginary square well
model throws a new light on some properties of the ${\cal PT}$
symmetric wave functions which are hardly accessible by
approximative techniques. In the nearest future, one can expect
that the further detailed study of the ${\cal PT}$ symmetric
square wells will give new answers to the recent puzzles as
formulated in ref. \cite{Sturm} and concerning the irregular
behaviour of the nodal zeros in the complex plane. Our present
example indicates that a surprising alternative to the Sturm
Liouville oscillation theorem could, perhaps, emerge in connection
with the study of zeros of the separate real and imaginary parts
of the ${\cal PT}$ symmetric wave functions.

\section*{Acknowledgement}

Inspiring communication with A. Mezincescu is gratefully
appreciated. Partially supported by the GA AS grant Nr. A 104
8004.

\section*{Figure captions}

\subsection*{Figure 1. Graphical
solution of eq. (\ref{enep}) ($y = \omega_N/2$, $T=1$)}

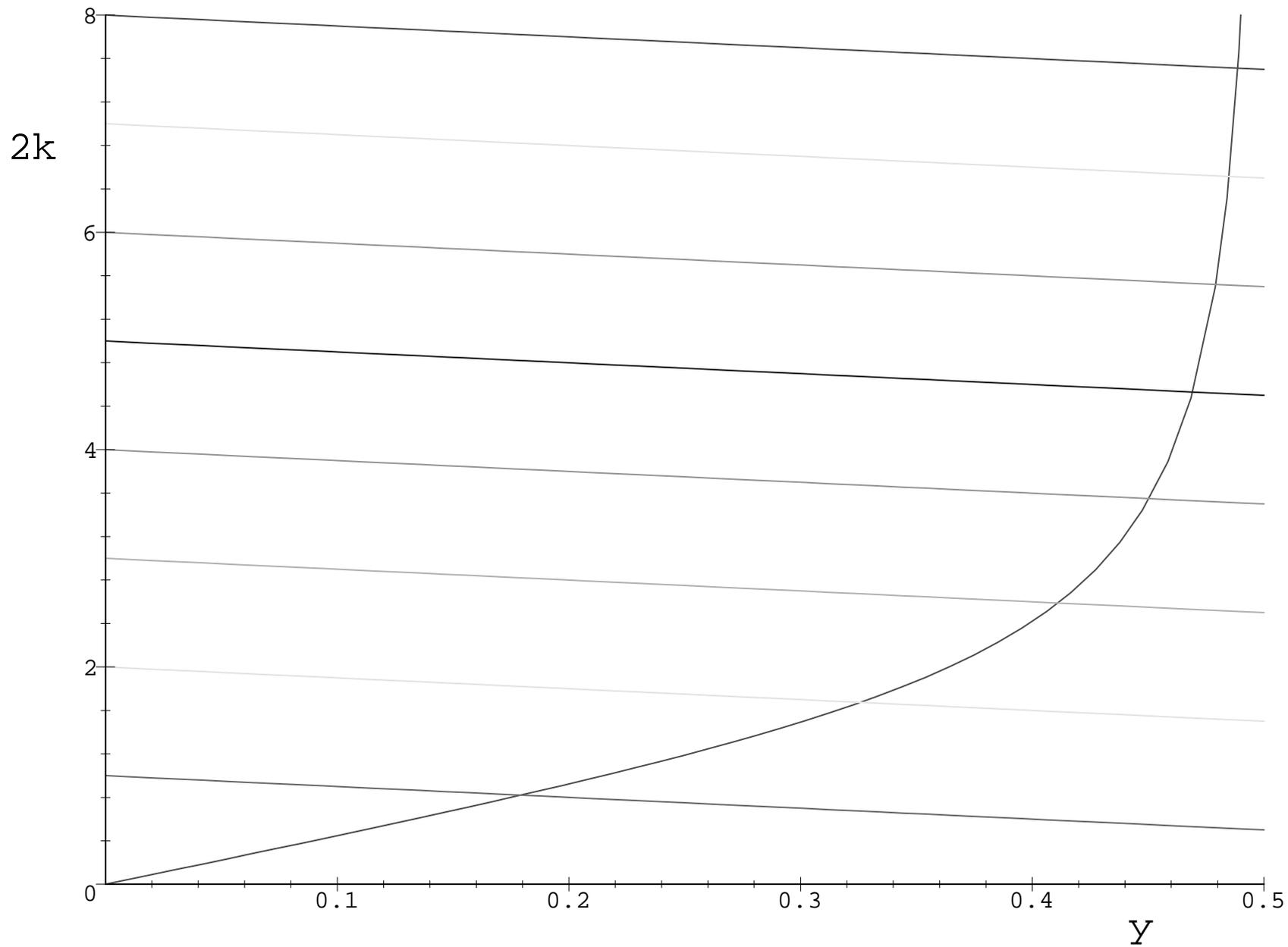
\end{document}